\documentclass[12pt]{iopart}

\usepackage{graphicx}  
\usepackage{dcolumn}   

\usepackage{color,soul}
\usepackage{float}
\usepackage{upgreek}
\usepackage[nolist,nohyperlinks]{acronym}
\usepackage{relsize}
\usepackage{xfrac}
\usepackage[latin1]{inputenc}
\usepackage[francais]{babel}
\usepackage{float}
\usepackage{subcaption}
\definecolor{blue}{rgb}{0.36, 0.54, 0.66}
\definecolor{amaranth}{rgb}{0.9, 0.17, 0.31}
\definecolor{pink}{rgb}{0.57, 0.36, 0.51}
\definecolor{ao}{rgb}{0.0, 0.5, 0.0}
\definecolor{maroon}{rgb}{0.76, 0.13, 0.28}
\definecolor{cardinal}{rgb}{0.77, 0.12, 0.23}

\usepackage[pagebackref=false,
			hyperindex=true,
			citecolor=cardinal,
			colorlinks=true,
			linkcolor=blue,
			urlcolor=blue]{hyperref}

\newcommand{\refeq}[1]{eq.~(\ref{#1})} 
\newcommand{\param}{\ensuremath{\vec{\theta}} }
\newcommand{\barD}{\overline{D}}
\newcommand{\barS}{\overline{S}}
\newcommand{\epp}{\epsilon^+}
\newcommand{\epm}{\epsilon^-}

\newcommand{\tildeD}{\tilde{D}}

\newcommand{\mrm}[1]{\mathrm{#1}}
\newcommand{\mcl}[1]{\mathcal{#1}}
\newcommand{\lb}{\left(}
\newcommand{\rb}{\right)}

\newcommand{\ba}{\begin{eqnarray}}
\newcommand{\ea}{\end{eqnarray}}
\newcommand{\be}{\begin{equation}}
\newcommand{\ee}{\end{equation}}
\newcommand{\bea}{\begin{eqnarray}}
\newcommand{\eea}{\end{eqnarray}}
\newcommand{\beq}{\begin{equation}}
\newcommand{\eeq}{\end{equation}}
\newcommand{\beqar}{\begin{eqnarray}}
\newcommand{\eeqar}{\end{eqnarray}}
\newcommand{\bc}{\begin{center}}
\newcommand{\ec}{\end{center}}
\newcommand{\noise}{\mathrm{n}}
\newcommand{\excess}{\mathrm{o}}

\begin{document}


\title[Assessing the detectability of a SGWB with LISA]{Assessing the detectability of a Stochastic Gravitational Wave Background with LISA, using an excess of power approach.}

\def\addressa{APC, Universit\'e Paris Diderot, CNRS/IN2P3, CEA/Irfu, Observatoire de Paris, Sorbonne Paris Cit\'e, 10 rue Alice Domont et L\'eonie Duquet, 75013 Paris, France}
\def\addressb{SYRTE, Observatoire de Paris, 61 Avenue de l'Observatoire, 75014 Paris}

\author{N.~Karnesis$^1$, M.~Lilley$^2$, A.~Petiteau$^1$}

\address{$^1$ \addressa}

\address{$^2$ \addressb}

\date{\today}

\begin{abstract}

The Laser Interferometer Space Antenna will be the first Gravitational Wave observatory in space. It is scheduled to fly in the early 2030's. LISA design predicts sensitivity levels that enable the detection a Stochastic Gravitational Wave Background signal. This stochastic type of signal is a superposition of signatures from sources that cannot be resolved individually and which are of various types, each one contributing with a different spectral shape. In this work we present a fast methodology to assess the detectability of a stationary, Gaussian, and isotropic stochastic signal in a set of frequency bins, combining information from the available data channels. We derive an analytic expression of the Bayes Factor between the instrumental noise-only and the signal plus instrumental noise models, that allows us to compute the detectability bounds of a given signal, as a function of frequency and prior knowledge on the instrumental noise spectrum.
\end{abstract}

\pacs{}
\maketitle

\section{Introduction
\label{section:introduction}}

The Laser Interferometer Space Antenna (LISA) is a space-borne Gravitational Wave (GW) observatory accepted by the European Space Agency (ESA) to be launched around 2034~\cite{2017arXiv170200786A}. LISA will be comprised of a constellation of three spacecraft forming an equilateral triangle with sides of 2.5 million kilometers. Each spacecraft will host two cubic test-masses maintained in free-fall conditions. The relative distance between the test-masses aboard the different spacecraft will be monitored by laser interferometry. LISA aims to directly measure GWs in the spectral range between 0.1 to 100~mHz. The predicted sensitivity of LISA extends the window for the detection of a Stochastic Gravitational Wave Background (SGWB) into the $\mathrm{mHz}$ band. The SGWB is a superposition of stochastic signals emitted from astrophysical and cosmological sources. The first type of source is the non-stationary, anisotropic, and partially unresolvable GW signal that is emitted by compact galactic binaries~\cite{Nissanke_2012}. This contribution is guaranteed to be detected by LISA, and will manifest itself as a confusion noise foreground that needs to be carefully handled in a data analysis pipeline. We also expect an extra component from Stellar Origin Black Hole Binaries,  as detected by LIGO-VIRGO observations~\cite{ligocatalogue}. This population of binaries is expected to contribute with a power law spectrum to the overall stochastic signal.  There is also the possibility to detect signatures originating from cosmological sources~\cite{Caprini_2018}, which would bring information on the properties of the primordial Universe and the physical processes describing its evolution. The main mechanisms behind the emission of cosmological stochastic signals are high-energy processes such as phase transitions, cosmic strings, and primordial black holes.

There exists an extensive amount of work describing methodologies to detect stochastic GW background signals, either with ground-based or with space-based detectors.  A review can be found in~\cite{sgwbunianalysis}. One of the most common approaches for the detection of this type of signal, is the cross-correlation of the outputs of a network of detectors~\cite{PhysRevD.46.5250, PhysRevD.79.062003}. This technique has been applied to data from ground-based detectors~\cite{Abbott_2007, PhysRevLett.118.121101, PhysRevD.79.062002}, and is based on constructing cross-correlation statistics as estimators of the energy density spectrum of a SGWB. Usually~\cite{PhysRevD.91.022003}, the search for spectral shapes is limited to a few representative cases (a flat spectrum for cosmological models for example). In the case of space-observatories and the isotropic component of a SGWB, one can search for and extract a particular spectral model in the data as in \cite{adamscornish1, adamscornish2}. In~\cite{adamscornish1}, the noise orthogonal TDI~\cite{tdi, aet} channels of LISA are used in order to distinguish between the instrument noise from a SGWB signal while the off-diagonal terms of the cross-spectrum matrix of the noise are used in order to constrain the parameters of the power spectrum of the instrumental noise. In~\cite{adamscornish2} the complexity was increased with the addition of an astrophysical confusion foreground signal in the data. In Pulsar Timing Arrays, upper limits have been set for isotropic SGWB in~\cite{Arzoumanian_2016, Lentati:2015qwp,PhysRevLett.115.041101,PhysRevD.79.084030} by using analysis techniques based on cross-correlation statistics.

Some work has dealt with the characterisation of the anisotropies of the stochastic signal~\cite{PhysRevD.92.102003, sgwbunianalysis, PhysRevD.71.024025, PhysRevLett.115.041101}. Anisotropies are investigated by searching for modulations of the correlated output of the detectors, or by constructing the sky maps of the strain intensity as a function of direction on the sky~\cite{PhysRevLett.122.081102}. The case of non-Gaussian~\cite{Seto_2008} signals present in Gaussian instrument noise has also been studied, and different approaches have been proposed for its characterization. For example, constraints on the SGWB signal can be estimated by employing higher-tail likelihood functions, as in~\cite{PhysRevD.89.124009}. On the other hand, generalised cross-correlation statistics~\cite{PhysRevD.89.124009} have been developed for the detection of signals in non-Gaussian noise.  At this point, it should be stressed that in all stochastic GW signal search algorithms, a sensible model of the instrument noise is essential. The method introduced in this paper also relies on properly modelled noise in order to extract the stochastic signal from the data.

The technique developed in the present work focuses on an analytic derivation of the detectability of an isotropic and Gaussian stochastic signal in the presence of Gaussian instrumental noise, with a particular application to LISA. We begin by assuming a noise power spectrum, with an uncertainty on its overall amplitude, parametrised by $\epsilon$. In this work, $\epsilon$ is independent of frequency, but this could easily be extended to the general case, thereby introducing an uncertainty on the noise spectral shape. Then, using the Gaussian properties of the signal and instrumental noise, we build a statistical framework for assessing signal detectability. We do so by determining the evidence for the presence of a signal in the data using the value of the Bayes factor, which we compute explicitly. Should our assumptions on the noise model fail, or should the parameter $\epsilon$ be underestimated, then the excess of power detected could be wrongly interpreted at a stochastic GW background instead of an unforeseen source of noise. On the contrary, if our noise model and its uncertainty as parameterised by $\epsilon$ is correct, then any excess power can be attributed to a GW signal with confidence. The approach introduced in this paper thus relies on the assumption of Gaussianity of both noise and signal, and on some level of knowledge of the noise, but is free of any assumption on the signal shape. This method was developed as a means to constrain the parameter space that generates detectable stochastic signals for LISA, without relying on specific stochastic GW signal spectral shape assumptions, or on time-consuming Monte Carlo simulations.

Our method expands on the technique employed in the data analysis of the LISA Pathfinder (LPF) mission~\cite{lpf_prl}, where one sought to characterise a noise contribution of unknown origin in the lower part of the differential acceleration spectrum between the two free-floating test masses~\cite{lpf_prl2}. Here, we first calculate numerically the PSD of the time series data, on a logarithmically equally-spaced frequency grid~\cite{trobsheinzel2006}. Then, we compute analytically the posterior probability of detection of a stochastic signal, as a function of a given instrument noise PSD shape and the parameter $\epsilon$. The binning strategy used in the computation of the numerical PSD allows us to consider each data point in the PSD as independent. As a result, we are able to estimate the stochastic signal power without the need for a signal PSD model. From there, we derive an analytic expression for the Bayes Factor between the {\it noise-only} and the {\it noise-plus-signal} models. The result of the present work is a tool that allows quick assessment of the detectability of a particular isotropic SGWB signal, given a certain level of confidence on the characteristics of the instrumental noise. However, it should be stressed that this methodology is built upon the assumption of ideal, Gaussian, stationary, and non-interrupted data, and is applied as proof-of-principle to simplified data sets that are free from ``loud'' transient signals. This being said, this approach can in fact be applied to more complicated cases, provided that the necessary statistical models of signal contaminations are considered in the analysis.

In section~\ref{section:theory} we discuss the theoretical approach used to model the power excess in the signal for each frequency bin.  The calculations presented in our work apply to an idealised LISA scenario, where the displacement noise ($S_\mrm{i}$) and the acceleration noise ($S_\mrm{a}$) are equal for all test-masses. This fact alone greatly simplifies the calculations, and allows us to derive analytic expressions for the signal detection statistics. In section~\ref{section:testcase} we apply our approach to the {\em Radler}~\cite{ldcdata} LISA Data Challenge (LDC) data set, as a simplified test case, but the application of our methodology to more complicated scenarios is also discussed. We then describe the detectability of a stochastic background as a function of its amplitude and the uncertainty on the noise power spectrum amplitude in section~\ref{section:detection}, by deriving an analytic expression for the Bayes factor. Finally, in section~\ref{section:discusssion} we summarise our main results and elaborate on the possible applications of this technique.

\section{Probability of power excess and application to synthetic test data
\label{section:theory}}

\subsection{Theoretical Background 
\label{section:onlytheory}}

Let us introduce the single channel data time series $d(t)$, which in our case is the time series of a single channel after applying the Time Delayed Interferometer (TDI)~\cite{tdi} algorithms. Then, and if we assume Gaussian and zero mean noise sources~\cite{rover2, stefanologarithmic}, the real and imaginary parts of the Fourier transform of the data $\tildeD$ at each Fourier coefficient with index $i$ are also independent Gaussian variables, provided that we correctly downsample the spectrum given the choice of windowing function used for its computation. In this case, the joint conditional probability density function for the power spectrum of the data, normalized by the corresponding theoretical power spectrum, follows a $\chi^2_k$ distribution with $k=2$ degrees of freedom~\cite{solomonpsd, Pieroni:2020rob}. This means that if we call $S_\mrm{d}$ the numerically computed power spectrum of the data, and $S_\mrm{t}$ their theoretical, or ``true" PSD, then $2S_\mrm{d}/S_\mrm{t}\sim\chi^2_2$. From that relation, we can derive that
\begin{equation}
	p(\tildeD| S_\mrm{t}) = \prod_{i} \frac{1}{ S_\mrm{t}[i]} \mrm{exp} \left( -\frac{S_\mrm{d}[i]}{S_\mrm{t} [i]} \right),
	\label{eq:post1}
\end{equation}
where $S_\mrm{d}[i]$ is taken at frequency $f[i]$, and where $S_\mrm{t} [i]$ is the model power spectrum at frequency $f[i]$, which, in the absence of spurious signals and other noises is the sum of a stochastic signal and the instrument noise. Finally, $i$ is the index running across the frequency grid. Under these ideal circumstances, we can safely split the data in $N$ segments and average them in frequency so that, in each bin $i$,
\begin{equation}
	p(\barD[i] | S_\mrm{t}) = \frac{ e^{ - \frac{ \sum_{j=1}^N D_j[i] }{S_\mrm{t}[i]} } } {S_\mrm{t}[i]^{N}} = \frac{ e^{ - N \frac{ \barD[i] }{S_\mrm{t}[i]} } } {S_\mrm{t}[i]^{N}},
	\label{eq:post2}
\end{equation}
where $\barD[i]$ is the average of the $N$ periodograms of the time series data in frequency bin $i$. Now, we can assume that the theoretical power $S_\mrm{t}[i]$ is the sum of the true signal plus the instrumental noise:
\begin{equation}
	S_\mrm{t}[i] = S_\mrm{o} [i] + S_\noise [i],
	\label{eq:noise}
\end{equation}
with $S_\noise [i]$ the instrumental noise, and $S_\mrm{o}$ the excess power measured for each frequency $f[i]$.

Let us also introduce an uncertainty in the noise amplitude by assigning a prior probability on the power spectrum level $S_\noise [i]$ per frequency bin $i$. We choose to use a uniform prior in each bin. This will also allow us to compute the integrals that follow analytically.  Defining the uncertainty in the noise amplitude by parameter $\epsilon[i]$, the instrument noise power spectrum lies in the range $[\barS_\noise [i] - \epsilon[i], \, \barS_\noise [i] + \epsilon [i]]$, where $\barS_\noise[i]$ is a best estimate, for each frequency $f[i]$. A step further would be to generalise to an asymmetric range around $\barS_\noise [i]$, which would then lead to $[\barS_\noise [i] - \epm[i], \, \barS_\noise [i] + \epp [i]]$. Marginalizing over $S_\noise$, the resulting PDF for each frequency bin is now
\begin{equation}
	p(\barD | S_\noise, S_\excess) = \int_{{\bar S}_\noise - \epm}^{{\bar S}_\noise + \epp}  \frac{ e^{ - N \frac{ \barD }{S_\excess+S_\noise} } }{\left( S_\excess + S_\noise \right)^{N}} \mrm{d} S_\noise .
	\label{eq:margpdf}
\end{equation}
where we have dropped the $[i]$ indices, for the sake of notational simplicity.
After a change of variable, we can use the incomplete gamma function $\Gamma_t (x) = \int_x^\infty y^{t-1} e^{y}\mrm{d}y$ in \refeq{eq:margpdf}. Then, the posterior PDF for the signal power $S_\excess [i]$ for each frequency $f[i]$ can be expressed as   
\begin{equation}
	p(S_\excess | \barD, S_\noise) = C \lb \Gamma_{N - 1} \lb A^+ \rb - \Gamma_{N - 1} \lb A^- \rb  \rb,
	\label{eq:post3}
\end{equation}
with
\be
	A^\pm = \frac{N \barD}{\barS_\noise + S_\excess \pm \epsilon^\pm} \qquad \text{and} \qquad
	C = \frac{1}{\lb \epp + \epm \rb \lb N \barD \rb^{N-1}} \,.
	\label{eq:constants}
\ee
Eq. (\ref{eq:post3}) is an analytic expression of the underlying stochastic signal for each frequency $f[i]$, given an uncertainty $\epsilon[i]$ of the instrumental noise. As we will see later in section (\ref{section:testcase}), it can be applied to simplified synthetic data to retrieve a first estimate of the properties of the signal.  

Let us now discuss the dependence of $S_\noise$ on a set of parameters $\param_n$. In this study, our simple noise model is a function of the test mass position and acceleration noise levels. Prior information on these parameters can be directly drawn from the main results of the LPF mission~\cite{lpf_prl,lpf_prl2}. In addition, it is expected that a first estimate of $\param_n$ from the onboard calibration measurements of the instrument will be obtained during the commissioning phase of the LISA mission.  In this work, we shall again assume that their prior densities follow $\param_\noise \sim U [\param_\noise^\mrm{min}, \param_\noise^\mrm{max}]$.  Consequently, the aforementioned limits generate lower and upper bounds on the overall power spectral density of the instrumental noise such that
\begin{equation}
	S_\noise^\mrm{min}[i] = S_\noise [i,\param_\noise^\mrm{min}] \leq S_\noise [i, \param_\noise] \leq S_\noise [i,\param_\noise^\mrm{max}] = S_\noise^\mrm{max}[i].
	\label{eq:limits}
\end{equation}
This translates into $S_\noise \in \barS_\noise [i] \pm \epsilon [i, \param_\noise]$. Thus, if we substitute \refeq{eq:noise} into (\ref{eq:post2}) and marginalise over $S_\noise$, the resulting PDF for each frequency with index $i$ is simply~\refeq{eq:margpdf}.

Having defined the statistical framework to be applied directly to the time series $d (t)$ in order to infer the signal $S_\excess [i]$, we now calculate the numerical power spectrum of the data. Following~\cite{trobsheinzel2006} and~\cite{heinzelfft}, we compute the power spectra of the time series on a downsampled logarithmic frequency axis. The power spectrum at each point $i$ on the frequency grid is calculated by taking into account the complete time-series data, and by adjusting the number of averages $N[i]$. In essence, short data segments are chosen for higher frequencies, while longer data segments are chosen for lower frequencies, yielding an accurate and improved estimation of the averaged spectra $\barD$ together with the associated errors in each bin\footnote{The power spectrum measurement errors for each frequency $f[i]$ can also be taken into account in the analysis, by absorbing them into $\epp[i]$ and $\epm[i]$.}. In~\cite{logpsdlpf, logpsd, lpf_prl2} the approach of~\cite{trobsheinzel2006} was extended so as to take into account the correlations between Fourier coefficients. This was achieved by carefully choosing $N[i]$ in a procedure that depends on the choice of windowing function and its spectral properties. Following the same approach, we generate a sparse grid frequency series with uncorrelated data points~\cite{lpf_prl2, stefanologarithmic}, allowing us to treat the data $\barD[i]$ independently. 

The main purpose of this work is to use of the analytic framework developed above to compute the detectability bounds and capabilities of LISA, for stationary, isotropic, and Gaussian SGWB signals (see section~\ref{section:detection} below).  This being said, \refeq{eq:post3} expresses the posterior probability of a GW signal power given the data, and we deem it useful to demonstrate the capabilities of this framework in an
illustrative parameter estimation application. In the following section, working with idealized data, we explore the behaviour of~\refeq{eq:post3}, and attempt to recover the overall shape of a stochastic GW signal. We then compare our findings with other parameter estimation techniques. In order to construct the posteriors of \refeq{eq:post3} for each frequency coefficient $i$, one starts with the calculation of the power spectra of the data $\barD$, using the numerical method of~\cite{trobsheinzel2006, lpf_prl2, logpsdlpf, logpsd}. The number of averages $N[i]$ per frequency coefficient $i$, which is an output of the numerical methods, is also substituted in~\refeq{eq:post3}. The joint posteriors over the data channels can then be mapped via grid methods. One can also use sampling Markov Chain Monte Carlo methods for more complicated cases, for example when the parameters of the noise model are to be estimated together with the signal parameters. In this work we perform both for cross-validation reasons, since the computational cost of doing so isn't too demanding. We use a Metropolis-Hastings Algorithm that is enhanced with a simulated annealing phase~\cite{Nofrarias2010, anneal2}, and adaptive proposal distribution mechanisms~\cite{Haario1999}. The sampler is then used to explore the posterior for the signal level (see \refeq{eq:post3}) for each of the frequencies of the analysis. Together with the noise model parameters, the dimensionality of the problem can reach the order of $10^2$ parameters.  

\subsection{Application to synthetic LISA data 
\label{section:testcase}}

We now apply the method described in Section~\ref{section:onlytheory} to LISA simulated data.  For our purposes, we choose to work with the {\em Radler} LISA Data Challenge data set~\cite{ldcdata} that contains only signals originating from stationary and isotropic stochastic GW sources. This particular dataset does not contain the expected astrophysical foregrounds, the anisotropic and non-stationary stochastic signals due to compact galactic binaries~\cite{Robson_2017}, or any other source of non-stationarity (i.e. noise transients, data gaps, etc). Indeed, the {\em Radler} data set does not represent a realistic scenario, but it can be used to apply the methodology described in the previous subsection~\ref{section:onlytheory} as a proof-of-principle. 

The data for all channels have been simulated using the same noise parameters.  This means that the displacement and acceleration noises for all test masses in all spacecraft are equal. In particular, the level of the acceleration noise for each test mass is $S_\mrm{a}= 3\times10^{-15} \mrm{m}^2/\mrm{sec}^4/\mrm{Hz}$, while the displacement noise level is $S_\mrm{i} = 15\times10^{-12} \mrm{m}^2/\mrm{Hz}$. In such a situation, it is convenient to work  with the noise orthogonal A, E, and T, TDI variables, expressed as linear combinations of the standard X, Y, and Z, TDI channels~\cite{tdi, aet}. This is because in the special case where all test-masses share common $S_\mrm{a}$ and $S_\mrm{i}$ noise levels, the cross-terms of the noise cross-spectrum matrix between the A, E, and T channels reduce to zero~\cite{adamscornish1}. Then, only the terms appearing on the diagonal of the matrix are to be considered, and from \refeq{eq:post3}, the posterior for a multi-channel analysis reduces to $\sum_{K}p(S_\excess | \barD_K, S_{\noise,\, K})$ for $K \in \{ \mrm{A},\, \mrm{E},\, \mrm{T}\}$.

\begin{figure}[ht!]
\bc
\includegraphics[width=0.6\textwidth]{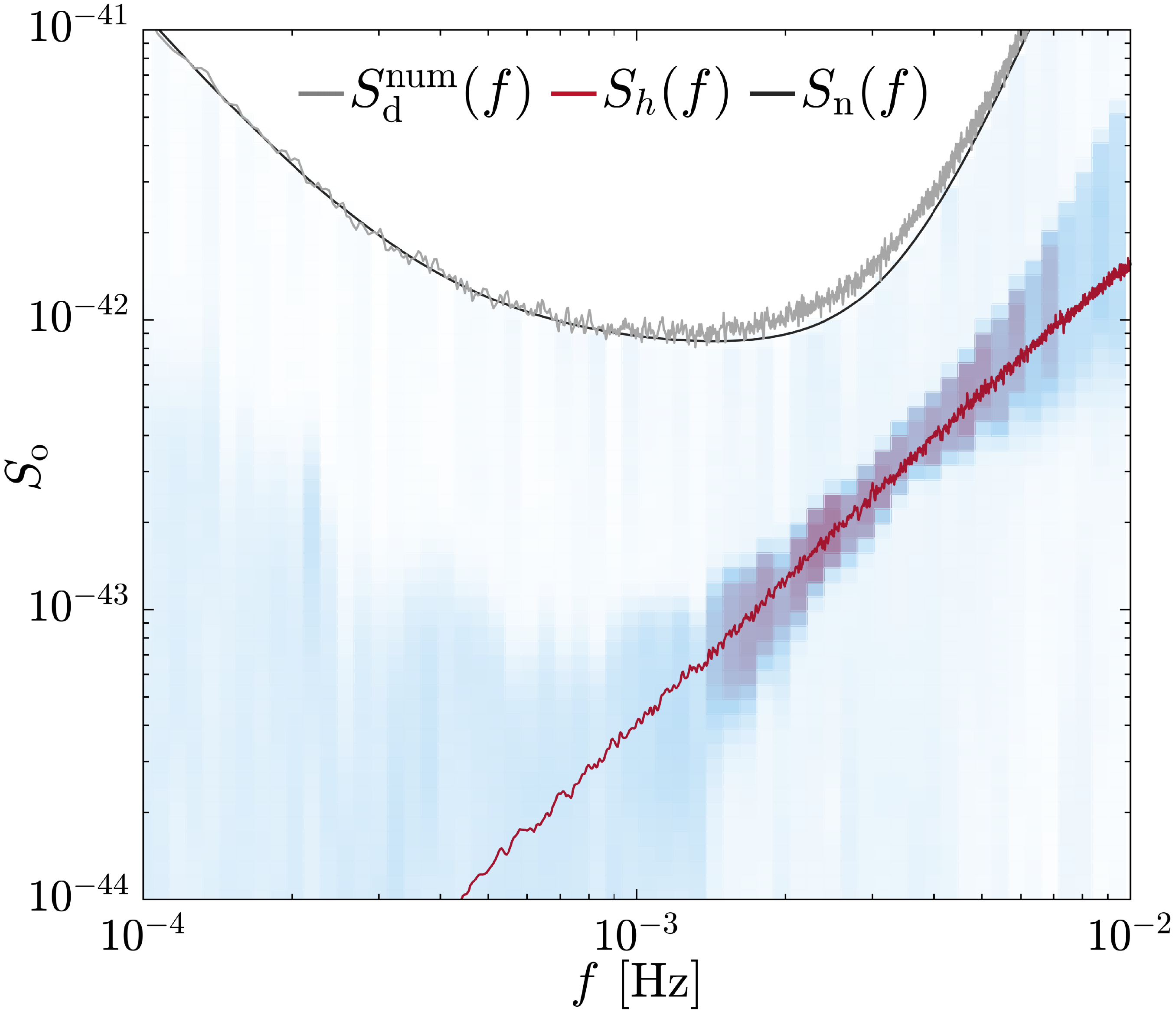}
\caption{The reconstruction of the SGWB signal for the {\em Radler} LDC dataset. The grey curve represents the numerically computed power spectrum $S_\mrm{d}^{\mrm{num}} (f)$ of the TDI A channel of the {\em Radler} data set, while the actual signal present in the data $S_h (f)$,  is depicted in red. The shaded area represents the normalized posterior probability densities for each of the spectral coefficients assumed in the analysis, and the PSD of the noise is shown in black. \label{fig:contourf}}
\ec
\end{figure}
The noise power spectra can be written as 
\ba
S_{\noise,\, \mrm{A}\mrm{E}}(f) = & 8\, \mrm{sin}(\lambda)^2 \big[ 2 S_\mrm{i, RF} \lb 3 + 2\mrm{cos}(\lambda)^2 + \mrm{cos}(2\lambda) \rb \nonumber\\
& + S_\mrm{a, RF} \lb 2+\mrm{cos}(\lambda)\rb  \big],
\label{eq:sim_psd_model}
\ea
where $\lambda = 2 \pi Lf/c $, $L$ is the LISA armlength \cite{ldcdoc}, while the noise parameters are expressed in relative frequency units, thus the 
subscript $\mrm{RF}$. The SGWB signal in the {\em Radler} dataset is a simple power law with amplitude $\mrm{log}_{10}(\Omega_0) = -8.445$ and spectral index $\alpha = 2/3$, where 
\begin{equation}
S_h = \frac{3H_0^2 \Omega_{\mrm{gw}} (f)}{4\pi^2 f^3}, \text{ with } 
\label{eq:sgwbsignal}
\end{equation}
\begin{equation}
\Omega_{\mrm{gw}} (f) = \Omega_0\left(\frac{f}{25 \mrm{Hz}}\right)^\alpha.
\label{eq:sgwbomega}
\end{equation}
In order to perform the analysis, one first computes the noise-orthogonal A, E and T variables, derived from X, Y and Z.  The data are then windowed, averaged in frequency, and binned. Together with the spectral coefficients that characterize the SGWB signal, the extra noise parameters $S_\mrm{a}$ and $S_\mrm{i}$ can also be sampled. The choice of prior for the noise power spectra is made using the relevant numbers from the results of LPF~\cite{lpf_prl,lpf_prl2}. In the frequency range of interest, the acceleration noise of each test-mass has an uncertainty of the order of $\sim1\%$. The error on the displacement noise for LISA is more difficult to establish, firstly because during the LPF mission only a local measurement was performed, and secondly due to the differences between the optical subsystems of LISA and LPF. Nevertheless, the technology to be used in LISA will be the same as that used in LPF, and the performance of the optical subsystems in the LPF mission was remarkable, almost two orders of magnitude better than requirements~\cite{lpf_prl}.  For LISA, it is therefore reasonable to expect an error on the displacement noise levels similar to that of LPF.  The confidence level on the overall power spectrum is thus chosen to be of the order of $\epp = \epm = \epsilon [i] = 0.05\, \barS_\noise [i]$ at all frequencies. Given the above, and for the purpose of this work, this value of $\epsilon[i]$, also being symmetric around $\barS_\noise [i]$, can be considered a conservative figure. 

The sampling of \refeq{eq:post3} yields the results presented in figure~\ref{fig:contourf}. The shaded area in this figure represents the normalized posterior probability densities for each frequency bin, while the red line indicates the true underlying signal that was injected in the data. 
The figure demonstrates that, for each frequency where the SNR of the signal is high enough, one can reconstruct the excessive signal due to the SGWB and separate it from the noise of the instrument. The reconstruction of the signal is also poorer, as expected, for lower signal-to-noise ratio areas of the spectrum. The parameters of \refeq{eq:sgwbsignal} and \refeq{eq:sgwbomega} can be extracted by fitting a line in log-space on the extracted data points, deconvolved by the LISA instrument response function $R(f)$, which is the response of LISA to a generic observed GW signal, and depends primarily on the orbits of the constellation. For a given LISA configuration, one can analytically approximate the response function~\cite{cutlerResponse, cornishResponse}, but here we directly use the LISA simulator \texttt{LISACode}~\cite{lisacode} to estimate it numerically. For the data presented in figure~\ref{fig:contourf}, we find 
\ba
\mrm{log}_{10}(\Omega_0) &=& -8.25\pm0.6\nonumber\\
\alpha &=& 0.72 \pm 0.2\nonumber
\ea
To put this result into perspective, we compare it with a direct model-based analysis (as in~\cite{adamscornish1, adamscornish2}) on the same data set. The model is a simple power law and the fit is performed at all frequencies.   Performing a joint fit of the SGWB spectral parameters and noise parameters, we obtain
\ba
\mrm{log}_{10}(\Omega_0) &=& -8.29\pm0.35\nonumber\\
\alpha &=& 0.68 \pm 0.09\nonumber
\ea
It can be seen that although the estimates made using both methods are within error of each other, the error bars are a factor 2 smaller in the model-based analysis\footnote{Note that in the model-based approach, the logarithm of the square of the acceleration and displacement noise parameters are estimated to be $\mrm{log}_{10}(S_\mrm{a}^2)=-29.042\pm0.002$ and $\mrm{log}_{10}(S_\mrm{i}^2)=-21.658\pm0.003$ respectively. The acceleration noise level agrees within 1-$\sigma$ with the true value (recall that $S_\mathrm{a}=3\times 10^{-15}\mathrm{m}^2/\mathrm{sec}^4/\mathrm{Hz}$), while the displacement noise is estimated to be within 3-$\sigma$ (recall that $S_\mrm{i}=15\times 10^{-12}\mathrm{m}^2/\mathrm{Hz}$).  This discrepancy is likely due to the difference of the analytical noise PSD of the TDI output with that produced by the simulator (see \refeq{eq:sim_psd_model}). Indeed, the simulator generates data in time domain, and includes additional effects such as anti-aliasing filters and interpolation in order to compute the TDI variables.}.

This being said, a direct comparison of the parameter estimation capabilities between the two methodologies is not straightforward. In the direct model-based approach, both the noise and SGWB model parameters are simultaneously fit across the entire frequency range.  The errors on those parameters are then obtained from their posterior distributions.  In contrast, the method developed in the present work first provides a bin-by-bin estimation of power excess caused by a stochastic signal, assuming, {\it a priori}, an uncertainty $\epsilon[i]$ on the level of noise in each bin.  Then, in order to do parameter estimation, one fits a power law, in log-space, on a downsampled frequency grid, on the excess of power signal extracted in step one.  It is thus reasonable to expect this level of disagreement on the error estimates of the parameters. Nevertheless, as already stated, the main point of this work is not to develop this methodology for parameter estimation, but rather as a tool to assess the detectability of any given stationary stochastic signal with LISA. This is the subject developed in the next section.

\section{Detectability assessment
\label{section:detection}}

Working in a Bayesian framework, model selection can be performed by
computing the ratio of the marginalized posterior distributions, or ``evidence'', between competing models. 
In this work, we are interested in developing a framework to assess the detectability of a given
stochastic GW signal. To do that, we need to calculate the Bayes Factor $\mcl B_{10}$~\cite{bf} for the two models of interest.  
The first model, $\mcl M_1$, corresponds to the model that supports the hypothesis of a stochastic signal being present in the data, 
while the second, $\mcl M_0$ corresponds to the instrumental noise only hypothesis. 
It's important to note that the models $\mcl M_1$ and $\mcl M_0$ are nested: $\mcl M_1$ 
reduces to the simpler $\mcl M_0$ for the parts of the spectrum where the signal is negligible 
(or $S_\excess \sim 0$). In that case, and because the priors of the parameters 
are uncorrelated, the Savage-Dickey (SD) density ratio can be calculated 
explicitly~\cite{SDDR1,SDDR2}. The SD ratio yields the Bayes factor 
by taking the ratio of the normalised marginal posterior on the additional parameter 
in $\mcl M_1$ over its prior, evaluated at the value of the parameter for which $\mcl M_1$ 
reduces to $\mcl M_0$. In our case this calculation is performed by combining~\refeq{eq:post3} 
and the prior on $S_\excess[i] \sim U[0, \,\kappa[i]]$. 
The SD density ratio is a very useful tool for computing the $\mcl B_{10}$ 
between nested models, but is prone to ineffective exploration of 
the parameter space~\cite{sddrdowns}. This can happen, for example, when the posterior 
shows little support at the true parameter value in the lower-dimension model. 
This does not apply for simplified cases as in the present analytic application (i.e. assume $\param_\noise$ constant), but more complicated scenarios require efficient sampling of 
the parameter space. Additionally, in order to calculate the SD ratio, one needs to first 
numerically normalise the posterior PDF, which adds an additional numerical 
step in the analysis. In the present work, however, having a relatively easily integrable 
expression for the posterior distribution allows us to analytically derive an 
expression for the $\mcl B_{10}$ directly. Given sufficient sampling exploration, 
the two approaches should yield the same result.

We begin by writing the evidence for $\mcl M_1$ as the double integral of \refeq{eq:post2} 
over the excess signal $S_\excess$ and noise $S_\noise$ for each frequency. For the case of the signal power $S_\excess[i]$ in each bin $i$, we choose to set a prior that follows a uniform distribution $U[0, \,\kappa[i]]$, for an arbitrary positive value of $\kappa[i]$, that will depend on the specifics of the particular investigation under study. One of the main reasons we choose a uniform prior, is that it enables us to carry out a relatively simple analytic derivation of the evidence. Then, once more dropping index $[i]$ for the sake of clarity, we write the evidence for model $\mcl M_1$ per frequency $f[i]$ as 
\be
p(\barD | \mcl M_1) = C \int_{\barS_\noise-\epm}^{\barS_\noise+\epp} \int_0^\kappa\frac{e^{-\frac{N\barD}{S_\excess+S_\noise}}}{\lb S_\excess+S_\noise \rb^N} {\mrm d} S_\excess {\mrm d} S_\noise ,
\label{eq:sn}
\ee
with $C$ a constant. Similarly, for the noise-only case $\mcl M_0$ we get
\be
P(\barD | \mcl M_0) = C' \int_{\barS_\noise-\epm}^{\barS_\noise+\epp} \frac{e^{-\frac{N\barD}{S_\noise}}}{S_\noise^N}{\mrm d}S_\noise\,.
\label{eq:no}
\ee 
Taking their ratio yields a Bayes factor that depends on the uncertainty $\epsilon$ of the power spectral 
density of the instrument noise $S_\noise$, and on the quantities $\barD$ and $N$ that are constant, and depend on the spectral preprocessing of the time series data. The Bayes factor is simply
\be
\mcl B_{10} (\epsilon) = \frac{P(\barD | \mcl M_1) }{P(\barD | \mcl M_0) }.
\label{eq:bf0}
\ee
To proceed with the calculation we first define $\alpha^\pm = \barS_\noise \pm \epsilon^\pm$ and $\beta^\pm = \barS_\noise \pm \epsilon^\pm + \kappa$. For the sake of convenience, we also define the following useful quantities:
\bea
\Gamma^{\alpha^\pm} &= \Gamma_{N-2}  \lb N\barD /  \alpha^\pm \rb, \\
\Gamma^{\beta^\pm} &= \Gamma_{N-2}  \lb N\barD /  \beta^\pm \rb.
\eea
Then, if we calculate \refeq{eq:bf0}, from \refeq{eq:no} and~(\ref{eq:sn}), and by taking into account that $\Gamma_n (x) = (n-1)\Gamma_{n-1}(x)$, we obtain
\bc
\bea
\mcl B_{10} (\epsilon) &=& \frac{\barD N \lb \Gamma^{\alpha^-} -\Gamma^{\beta^-} -\Gamma^{\alpha^+} + \Gamma^{\beta^+} \rb }{\kappa(N-2)\lb \Gamma^{\alpha^-} - \Gamma^{\alpha^+}\rb}\nonumber\\
& &+ \frac{ \lb  \beta^-\Gamma^{\beta^-} - \alpha^- \Gamma^{\alpha^-} + \alpha^+\Gamma^{\alpha^+} - \beta^+\Gamma^{\beta^+} \rb }{\kappa\lb \Gamma^{\alpha^-} - \Gamma^{\alpha^+}\rb},
\label{eq:bf}
\eea
\ec
where again we have omitted the $[i]$ indices for the sake of clarity. The above expression holds for each frequency $f[i]$, and provides an estimate of the Bayes Factor between the two models of interest at $f[i]$.  Eq.~(\ref{eq:bf}) can be therefore be used to assess the detectability of a stochastic gravitational wave signal as a function of the level of uncertainty in the noise spectrum on the given grid of frequencies for the particular data set $\barD$. 
In the following sections, we find that the SD density ratio yields results identical to \refeq{eq:bf}. This provides cross-validation for \refeq{eq:bf}, and also demonstrates that the SD density ratio approximation to the Bayes Factor can be used as well.

In Section \ref{section:bfdata}, we apply our findings to the {\em Radler} data set, and in \ref{section:bfassess} we will use it to introduce a model-independent framework that will allow us to assess the detectability of isotropic stochastic GW signals.

\subsection{Application to synthetic data: Detectability assessment as a function of $\epsilon$}
\label{section:bfdata}

\begin{figure}[H]
\centering
\subcaptionbox{\label{fig:epsilonradler}}
{\includegraphics[width=0.48\textwidth]{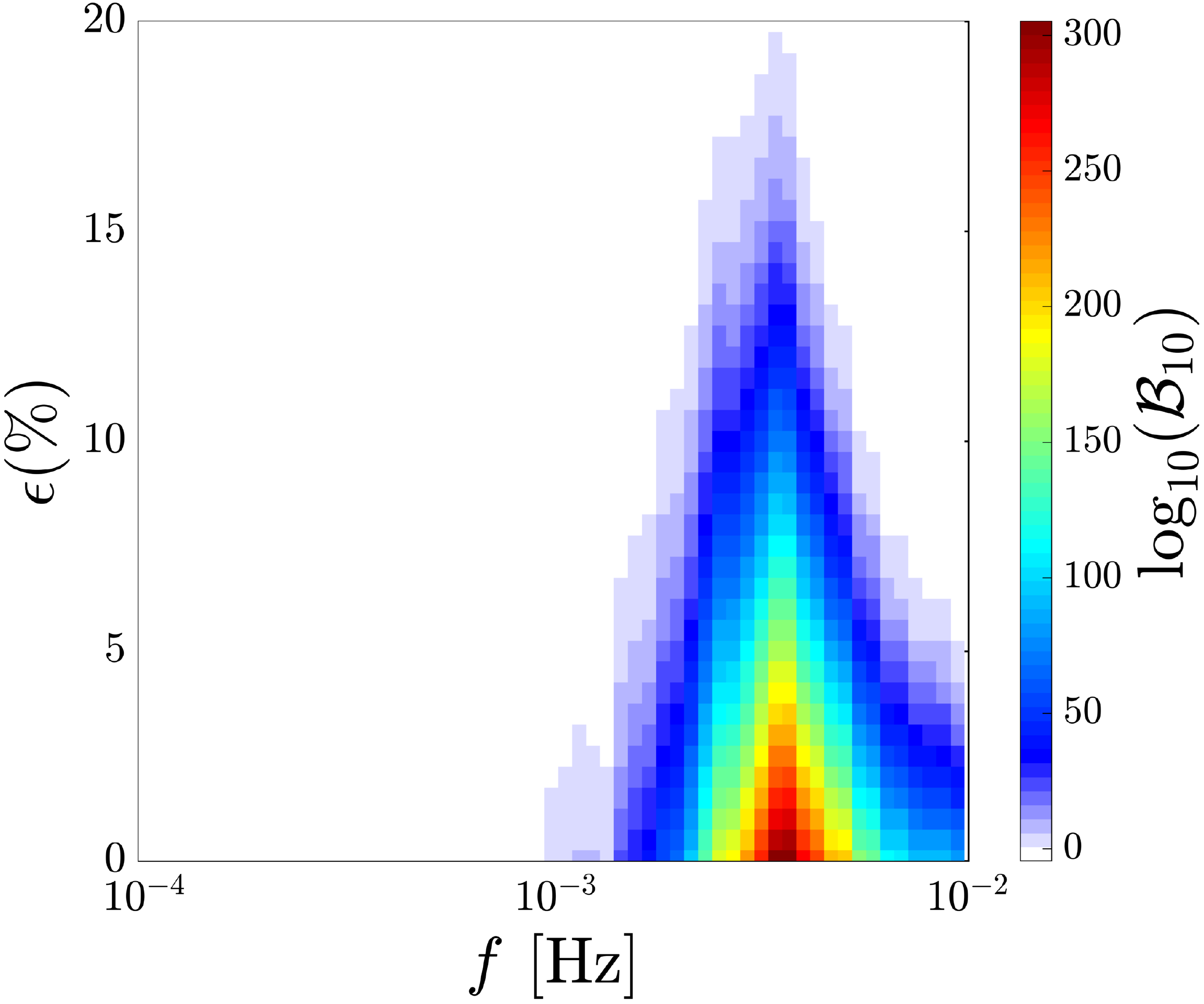}}
\subcaptionbox{\label{fig:detectability}}
{\includegraphics[width=0.46\textwidth]{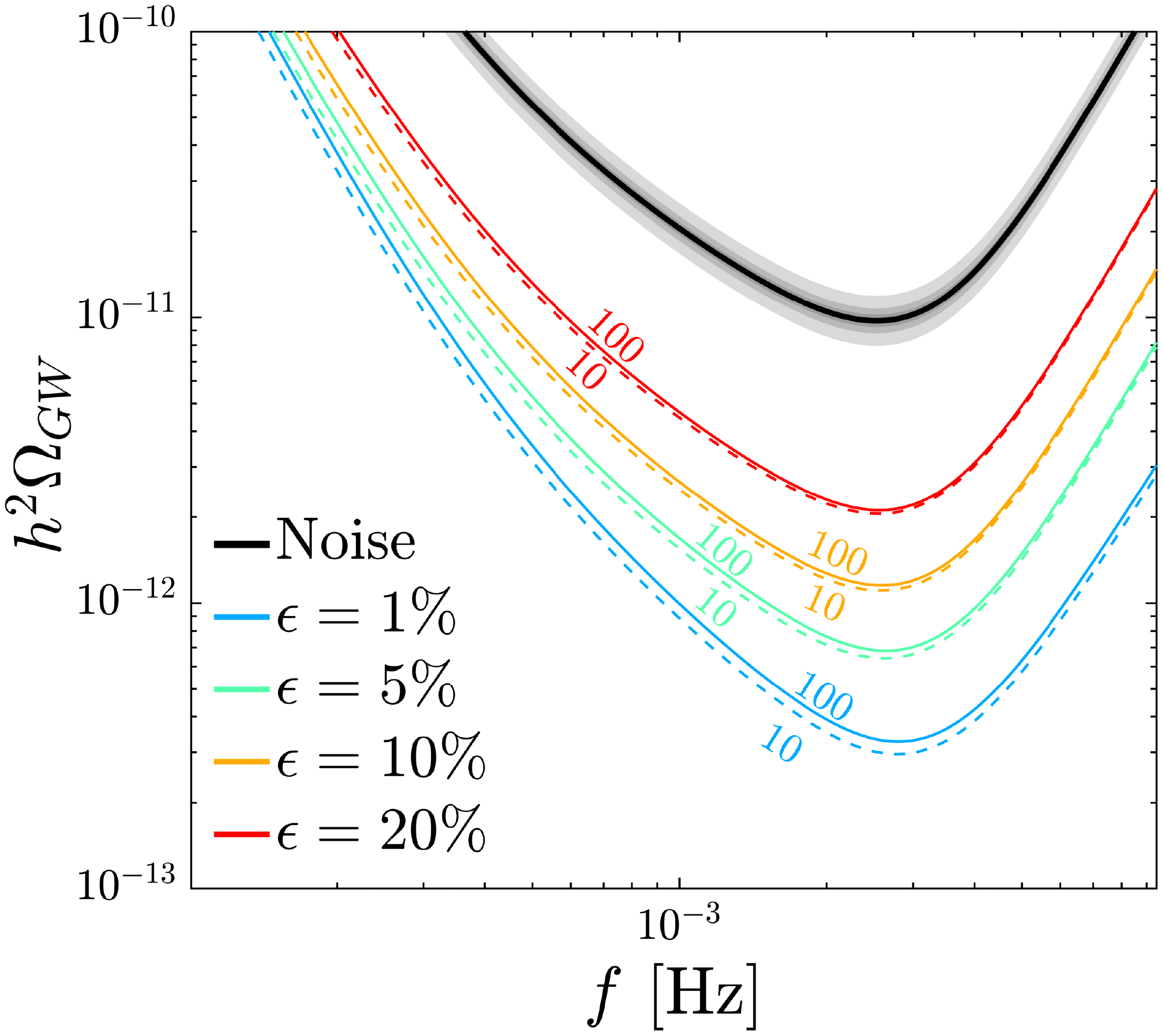}}
\caption{\ref{fig:epsilonradler}: The logarithm of the Bayes factor $\mcl B_{10}$ as a function of the relative uncertainty in the instrumental noise spectrum, $\epsilon$, comparing models  $\mcl M_0$ and $\mcl M_1$ applied to  the {\em Radler} LDC data set. The first model $\mcl M_1$, assumes the presence of GW signal in the data, while $\mcl M_0$ corresponds to the instrumental noise only case. As shown in figure~\ref{fig:contourf}, high SNR signals are obtained for frequencies above $1$~mHz, so that the greater values of $\mcl B_{10}$  are obtained for these particular frequencies. \ref{fig:detectability}: The values of the Bayes factor $\mcl B_{10}$ as a function of the stochastic signal power at each frequency of the analysis. The dashed lines are the contour lines for $\mcl B_{10}=10$ for $\epsilon = 1\%$, $5\%$, $10\%$, and $20\%$, while the solid lines correspond to $\mcl B_{10}=100$. The two levels of $\mcl B_{10}$ correspond, respectively, to positive and very strong evidence of a SGWB signal present in the data. The black solid line represents the corresponding noise level. This figure has been produced for a data set similar to the {\em Radler} dataset, i.e. assuming an uninterrupted data series of length of two years for all TDI channels (see text for details). With this strategy one can directly derive the detectability capabilities of LISA given different stochastic signals. The given signal components with amplitudes higher than the solid lines in this plot, would be detected by LISA (in this particular configuration), with increased confidence.}\label{fig:assessdetect}
\end{figure}

We now apply \refeq{eq:bf} to the {\it Radler} data set, and explore the detectability of the signal, as a function of the knowledge we have on the amplitude of the noise. We have already computed all quantities necessary to the calculation of~\refeq{eq:bf} in the analysis described in section~\ref{section:testcase}, namely the power spectrum of the data $\barD$, and the number of averages $N[i]$ per frequency $f[i]$. Then, for each frequency $f[i]$ we compute \refeq{eq:bf} for different values of $\epsilon$. For the case of the signal power $S_\mathrm{o}[i]$ at each bin $i$, we choose to set a prior that follows a uniform distribution $U[0, \kappa[i]\equiv\barS_\noise[i]]$, assuming that the power spectrum of the signal that we are interested in is lower at all frequencies than the spectrum of the instrument noise. This prior is suitable for our purposes as we are interested in assessing the detectability of weak signals, dominated by the noise. The result is shown in figure~\ref{fig:epsilonradler}, where the logarithm of $\mcl B_{10}$ is plotted for each frequency bin as a function of the value of $\epsilon$. As expected, for frequencies lower than $\sim1$~mHz where the signal-to-noise ratio is low, the $\mcl B_{10}$ takes small values, lending support to the noise-only model $\mcl M_0$. Figure~\ref{fig:epsilonradler} also demonstrates that for this idealized dataset, as expected, the detection of a SGWB with $\mrm{log}_{10}(\Omega_0) = -8.445$ and $\alpha = 2/3$ is becoming more challenging when our knowledge of the instrumental noise reaches the quite pessimistic 20\% margin, for each of the available data channels. 

\subsection{A methodology to assess the detectability of generic stochastic signals   
\label{section:bfassess}}
 
Reversing the strategy of Section~\ref{section:bfdata}, one can instead consider a fixed value for $\epsilon$ for all frequencies, 
and predict the gravitational wave signal amplitudes that would give values of $\mcl B_{10}$ providing positive evidence of model $\mcl M_1$, given the data. This provides a probability to detect a stochastic signal independent of the underlying physical model. To do so, let us begin by assuming a measurement with the same properties as the one of the {\it Radler} dataset. The hypothesized dataset therefore consists in an idealized and uninterrupted data stream, with Gaussian and mean zero noise, for a duration of two years. We then substitute $\barD$ with $\barS_\noise + S_\excess$ and compute the logarithm of $\mcl B_{10}$ given by \refeq{eq:bf}, for different values of the stochastic signal level $S_\excess$. We then infer the signal $S_\excess$ levels that would enable detection, given a particular instrument configuration, observation duration, and confidence in the instrument noise power spectrum for each frequency bin. 

The results of this computation are shown in figure~\ref{fig:detectability}.  The dashed lines correspond to the contour levels for $\mcl B_{10}=10$, while the solid lines correspond to values of Bayes factors of $\mcl B_{10}=100$. These values of the Bayes factor correspond, respectively, to positive and very strong~\cite{KassRaftery} evidence for the presence of a stochastic signal in the data. The calculation is repeated for $\epsilon = 1\%,\, 5\%$, $10\%$ and $20\%$, represented in figure~\ref{fig:detectability} in blue, green, yellow, and red, respectively. As expected, when the level of uncertainty in the noise increases, the stochastic signal amplitude required for detection increases. This methodology can be employed to assess the detectability capabilities of LISA for any stochastic signal at a given set of frequencies. Indeed, as already described, given the set of frequencies $f[i]$ and the values of $N[i]$ as derived from numerical power spectrum estimation methodologies~\cite{logpsdlpf, logpsd, lpf_prl2}, a theoretical power spectrum $S_\excess[i]$ as predicted by a given cosmological model, and finally the instrumental noise model with a chosen uncertainty $\epsilon$, \refeq{eq:bf} can be computed for all frequencies. The detectability of the particular model can be assessed both per frequency, and by summing over frequencies, $\sum_f \mathrm{log}B_{10}(f)$, as the data at each value of $f[i]$ are treated as independent measurements.

\section{Discussion \label{section:discusssion}}

We have presented a methodology to assess the detectability of an underlying stationary stochastic signal from the LISA TDI measurements which takes into account the uncertainty in the instrumental noise. We have expanded from a technique that had been employed to analyze the LPF data~\cite{lpf_prl2}. In LPF, this technique was utilized to identify excess noise power of unknown origin in the lower frequency part of the differential acceleration spectra, measured throughout the mission.  Here, we have adapted this framework in order to perform two tasks. We first applied it to a simulated dataset in order to characterize its stochastic signal, also taking into account a degree of uncertainty on the detector's instrumental noise. Secondly, we used this framework in order to predict the detectability of any given isotropic stochastic GW signal with LISA, in the case of idealised scenarios. This provides the possibility to study SGWB for various cosmological models by investigating the regions in parameter space that would yield stochastic GW backgrounds detectable with LISA. This approach is a very efficient alternative to Monte Carlo simulations.  In order to achieve the same goal using Monte Carlo simulations, one would need to recompute the marginal likelihood for multiple noise realisations.

In the first part of this work, the methodology was tested using the {\em Radler} simulated dataset provided by the LISA Data Challenge. This dataset was generated with a simplified LISA detector, with stationary Gaussian noise without bright spurious signals and with acceleration noise $S_\mrm{a}$ and displacement noise $S_\mrm{i}$ equal for all test-masses. The data only contains a stationary stochastic Gravitational-Wave background signal that follows a power law.  The fact that $S_\mrm{i}$ and $S_\mrm{a}$ are equal for all test-masses implies that, when working with the noise orthogonal A, E, and T TDI variables, the off-diagonal terms of the cross-spectrum matrix of the noise reduce to zero. Thus, only the diagonal elements of the cross-spectrum noise matrix need be considered in the analysis, and the total posterior reduces to the sum of posterior densities over the individual A, E, and T channels. 

This being said, the methodology can easily be extended to the more complicated case in which each of the readout channels have different noise properties. By including the cross-power spectra in the analysis one should be able to recover combinations of the instrument noise parameters~\cite{adamscornish1}. As expected, the GW spectrum is best recovered in the frequency range where the signal-to-noise ratio is high (see figure~\ref{fig:contourf}). The estimated data points of $S_\excess$ are then used in a power law fit in order to infer the slope and amplitude parameters of the underlying SGWB model (see \refeq{eq:sgwbsignal}). Such a fit yields parameter estimates that are in agreement with other methodologies, but with a disagreement on the recovered error-bars (see section~\ref{section:testcase}). This being said, as explained in Section \ref{section:testcase}, a direct comparison between the hierarchical method presented here and a direct model-based approach is not straightforward. In order for the two to be compatible, one would need to first consider the same prior ranges on the noise model. Secondly, the same data treatment would be required, i.e. same methods of power spectra estimates.

The main point of this work was not to develop this framework for parameter estimation, but rather to build a tool to assess the detectability of any given stationary stochastic signal with LISA, assuming ideal data quality. The approach we propose indeed provides an analytic and model-independent first level characterisation of any given underlying stationary power excess, as measured by different channels simultaneously.  We first integrate the expression of the posterior densities for the stochastic signal over the parameter space in two cases: data plus signal $\mcl M_1$, and instrument noise only, $\mcl M_0$. We then compute the ratio, which is an analytic formula for the Bayes factor $\mcl B_{10}$ between those two models, that depends on the level of uncertainty of the noise power spectral density. This expression (see~\refeq{eq:bf}) can directly be used to assess the efficiency of LISA at detecting a SGWB, given the data, depending on the instrument noise uncertainty at each frequency. 

In the case of the {\em Radler} dataset, we found that the high SNR signal in parts of the frequency range can be detected with increased confidence when the noise PSD for each data channel is less than $\sim$20\% (see figure~\ref{fig:epsilonradler}). In addition, reversing the argument, we studied the efficiency of LISA in detecting various levels of stochastic signals, 
given a particular uncertainty on the instrument noise PSD and observation time.  As expected, the SGWB signal that yields a positive $\mcl B_{10}$ greatly depends on our knowledge of the noise spectrum $S_\noise$.  In this work we used a level of noise uncertainty parameterized by constant bounds $\epm(f)=\epm$  and $\epp(f)=\epp$ over the frequency range of interest, but one could of course consider a more realistic case, by taking into account different level of uncertainties in different parts of the frequency range. The magnitude of $\epsilon(f)$ can depend, for example, on known instrumental noise features, like noise transients, noise bursts and other non-stationarities, on the imperfect subtraction of brighter sources, or finally from imperfect knowledge of foreground noises, such as the one produced by the compact galactic binaries. 

It is important to note, however, that this method cannot disentangle the various contributions to the overall signal and does not provide any information about the possible sources of the excess of power detected. Indeed, while this technique can easily be applied in the case of a simplified scenario (as is the case of the {\em Radler} dataset), applying it to a more realistic situation would first require that one take into account all the different contributions to the overall power spectrum. Any type of unidentified power in the data, such as noise non-stationarity, the presence of residuals, or any sort of imperfection, would distort the final result if not properly considered in the analysis. 

Finally we wish to comment on recent work of~\cite{binning} on the same topic. In~\cite{binning}, the observed power spectra are segmented to bins with  arbitrary initial widths. The data model for each bin is parameterized by an amplitude, a spectral slope and a noise model. Model parameters are then determined numerically assuming a Gaussian likelihood function, and by employing maximization algorithms. Then, model selection criteria are used in order to make a decision about joining adjacent bins, thus improving the parameter estimation step. This is an iterative and adaptive procedure capable of determining the spectral shape and amplitude of stochastic backgrounds without the need for a single model. In contrast, our main result derives from \refeq{eq:post3}, which yields an analytic estimate of the signal amplitude's posterior probability distribution on a sparse and equally spaced in log-space frequency grid. As expected, comparison between the two methods is not straightforward (see section \ref{section:testcase}). Starting from~\refeq{eq:post3}, we can derive an analytic framework to assess the detectability of any given Gaussian stationary stochastic signal. We do that by calculating the analytic expression for the Bayes factor between the noise plus signal $\mcl M_0$ and the noise-only $\mcl M_1$ models (see section \ref{section:detection}). This is particularly useful if one wishes to gain intuition on the efficiency of LISA in detecting SGWBs given a certain level of noise uncertainty, without the need for any simulated data, or computationally expensive Monte Carlo simulations. 


\ack
The authors would like to acknowledge the work of the LISA Data Challenges Working Group of the LISA Consortium, for the generation of the {\em Radler} data set. NK would like to thank the CNES DIA-PF post-doctoral program.


\medskip
\section*{References}

\smallskip

\end{document}